\documentclass{jkas}

\def\year{2015} 
\def\month{April} 
\def\volume{48} 
\def\issue{2} 
\def\beginpage{113} 
\def\endpage{123} 
\def\received{December 24, 2014} 
\def\accepted{March 16, 2015} 
\date{Received \received; accepted \accepted}

\usepackage{flushend} 
\newcommand{\farcm}{\mbox{\ensuremath{.\mkern-4mu^\prime}}}
\newcommand{\farcs}{\mbox{\ensuremath{.\!\!^{\prime\prime}}}}
\newcommand{\fdg}{\mbox{\ensuremath{.\!\!^\circ}}}
\newcommand\ion[2]{#1$\;${\scshape{#2}}}

\newcommand{\FeII}{[\ion{Fe}{ii}]}
\newcommand{\SII}{[\ion{S}{ii}]}
\newcommand{\OI}{[\ion{O}{i}]}
\newcommand{\NII}{[\ion{N}{ii}]}

\newcommand{\kms}{km~s$^{-1}$}

\newcommand{\degree}{^{\circ}}
\newcommand{\apj}{ApJ}                                       

\newcommand{\Ha}{H${\alpha}$}
\newcommand{\arcsec}{^{\prime\prime}}

\title{
Long-Slit Spectroscopy of Parsec-Scale Jets from DG Tauri
}

\author[1,2]{Heeyoung~Oh}
\author[3,4]{Tae-Soo~Pyo}
\author[2]{In-Soo~Yuk}
\author[1,2]{Byeong-Gon~Park}

\affil[1]{Korea University of Science and Technology, 217 Gajeong-ro, Yuseong-gu, Daejeon 305-350, Korea \email{hyoh@kasi.re.kr,~bgpark@kasi.re.kr}}
\affil[2]{Korea Astronomy and Space Science Institute,776 Daedeokdae-ro, Yuseong-gu, Daejeon 305-348, Korea \email{yukis@kasi.re.kr}}
\affil[3]{Subaru Telescope, National Astronomical Observatory of Japan, 650 North A'ohoku Place, Hilo, HI 96720, USA \email{pyo@subaru.naoj.org}}
\affil[4]{School of Mathematical and Physical Science, SOKENDAI (The Graduate University for Advanced Studies), Hayama, Kanagawa 240-0193, Japan}

\def\corrauthor{
T.-S. Pyo
}

\def\runningauthor{
Oh et al.
}

\def\runningtitle{
Long-Slit Spectroscopy of Parsec-Scale Jets from DG Tauri
}

\def\keywords{
stars: formation --- stars: outflows --- Herbig-Haro obejcts --- stars: individual: DG Tau --- techniques: spectroscopy
}

\def\abstracttext{
We present observational results from optical long-slit spectroscopy of parsec-scale jets of DG Tau. From HH 158 and HH 702, the optical emission lines of \Ha, \OI~$\lambda\lambda$6300, 6363, \NII~$\lambda\lambda$6548, 6584, and \SII~$\lambda\lambda$6716, 6731 are obtained. The kinematics and physical properties (i.e., electron density, electron temperature, ionization fraction, and mass-loss rate) are investigated along the blueshifted jet up to 650$\arcsec$ distance from the source. For HH 158, the radial velocity ranges from $-$50 to $-$250 km s$^{-1}$. The proper motion of the knots is 0{\farcs}196 $-$  0{\farcs}272 yr$^{-1}$. The electron density is $\sim$10$^{4}$ cm$^{-3}$ close to the star, and decreases to $\sim$10$^{2}$ cm$^{-3}$ at 14$\arcsec$ away from the star. Ionization fraction indicates that the gas is almost neutral in the vicinity of the source. It increases up to over 0.4 along the distance. HH 702 is located at ~650$\arcsec$ from the source. It shows $\sim$ $-$80 km s$^{-1}$ in the radial velocity.  Its line ratios are similar to those at knot C of HH 158. The mass-loss rate is estimated to be about $\sim$ 10$^{-7}$ $M_{\odot}$ yr$^{-1}$, which is similar to values obtained from previous studies.
}

\begin{document}
\jkashead 


\section{Introduction\label{sec:intro}}

In the formation of stars and planetary systems, studying outflows and mass-accretion is essential \citep{Hartigan1995}. Through the analysis of the kinematics and physical conditions of outflows from young stellar objects (YSOs), we are able to understand the jet launching mechanism and the interaction with ambient material. Parsec-scale jets are useful to study star formation in somewhat long time scale and large spatial distribution. There are a number of parsec-scale jets known \citep[e.g.,][]{Ray1987,Bally1994,Ogura1995,Bally1995}. One of the extreme cases is the HH 222 system, suggested as a giant Herbig-Haro flow spread over 5.3 parsec in length \citep{Reipurth2013}.

\citet{Mundt1983} discovered a jet-like outflow from DG Tau. DG Tau is an early evolutional stage CTTS (Class II) with mass of 0.67 $M_{\odot}$ \citep{Hartigan1995} and age of about 3 $\times$ 10$^{\rm 5}$ yr \citep{Beckwith1990}. \citet{McGroarty2004} suggested that it is the driving source of a parsec-scale jet including HH 158, HH 702, and HH 830. The later two HH objects are located at more than 9$^{\prime}$ away from DG Tau. In their follow up observations, \citet{McGroarty2007} reported that HH 702 showed receding motion with respect to DG Tau. They also concluded that HH 830 is not ejected from DG Tau. \citet{Eisloffel1998} showed that the HH 158 outflow extends to $\sim$ 11$\arcsec$ toward the southwest direction and estimated the direction of the proper motion of knots from DG Tau as $\sim$226 $\pm$ 7$\degree$. They also calculated the jet inclination angle as $\sim$ 32$\degree$ with respect to the line of sight. In addition to an optical jet, \citet{Rodriguez2012} and \citet{Lynch2013} detected a radio knot at $\sim$ 7$\arcsec$ from the central source. High spatial resolution observations with a space telescope and ground based large telescope with an adaptive optics system showed that the blueshifted HH 158 jet can be traced to within $\sim$ 0\farcs1 from the star \citep{Bacciotti2000, Bacciotti2002, Pyo2003, Maurri2014}. \citet{Lavalley1997} reported that the redshifted jet with the length of $\sim$ 1$\arcsec$. \citet{Pyo2003} showed that there is a 0\farcs7 gap between the redshifted jet and the star in the \FeII~1.644 $\mu$m emission, which indicates the presence of an optically thick disk. The two velocity components are shown in the optical and near infrared forbidden emission lines \citep{Hartmann1989, Solf1993, Hamann1994, Hartigan1995, Pyo2003, Pyo2006}.

In this paper, we report the results of optical long-slit spectroscopy on HH 158 and HH 702 outflows. We investigate the electron density, electron temperature, and ionization fraction from close vicinity of the central source to the knot at 14$\arcsec$ of HH 158 jet and HH 702 at 650$\arcsec$ from DG Tau. We note that this is the first time to obtain optical spectroscopy of the HH 702 outflow. We also summarize the proper motions of the knots in the jet with the positions during the last 30 years.
In Section 2, we describe the observation and data reduction. In Section 3, we show the line intensity and ratios in the position-velocity map. Sections 4 and 5 are the discussion and summary, respectively.
\begin{table*}[t]
\caption{Log of Observations.\label{tab:jkastable1}}
\centering
\begin{tabular}{ccccccc}
\toprule
 &  &  & Exposure & Slit  & Spectral & Spectral \\
Object& Date (UT)& P.A. ($\degree$)& time (s) & width ($^{\prime\prime}$) & resolution ($\lambda/\Delta\lambda$) & coverage (\AA) \\
\midrule
HH 158 & 2012 Nov. 15 & 223 & 1800 & 2.9 & 2070 & 5900--7050 \\
HH 702 & 2012 Nov. 15 & 193 & 2800 & 2.9 & 2070 & 5900--7050 \\
\bottomrule
\end{tabular}
\end{table*}
\begin{figure}
\centering
\includegraphics[trim=0mm 0mm 3mm 0mm, clip, width=80mm]{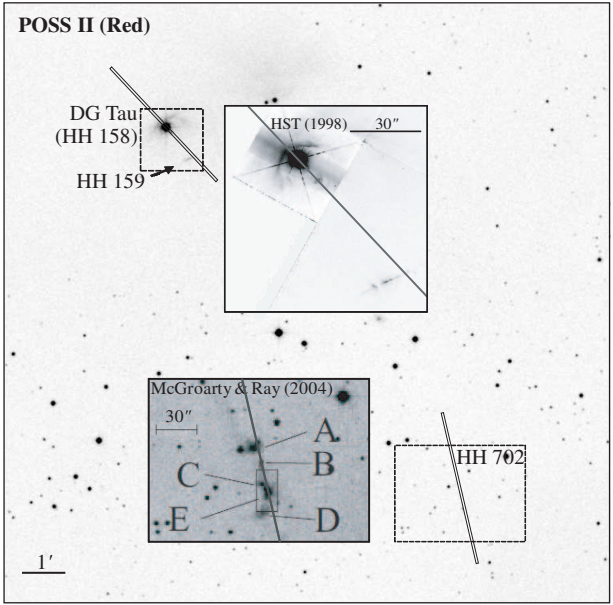}
\caption{Slit positions on the object field. The center of the 2\farcs9(W) $\times$ 3\farcm6(L) slit is set on HH 158 and HH 702. The position angles are 223$\degree$ and 193$\degree$ for HH 158 and HH 702, respectively. The background image is from the Second Palomar Observatory Sky Survey (POSS II). The HST image of HH 158, 159 and {\Ha} image of  HH 702 from \citet{McGroarty2004} are inserted as insets. Knot A $-$ E of HH 702 are marked. \label{fig:jkasfig1}}
\end{figure}
\begin{figure}
\centering
\includegraphics[width=83mm]{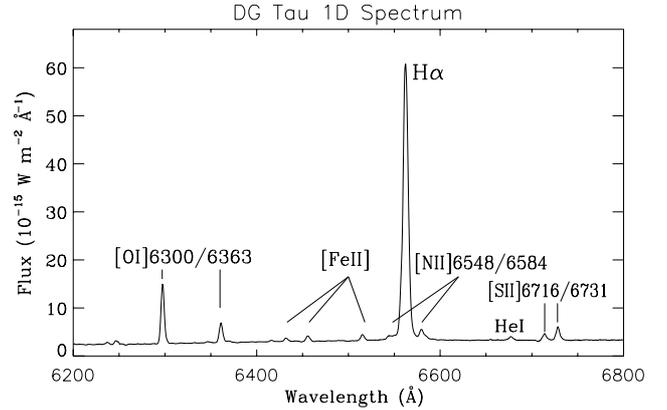}
\caption{Integrated 1D spectrum of the central region of DG Tau. \label{fig:jkasfig2}}
\end{figure}

\section{Observation and Data Reduction\label{sec:Observation}}
The observation was conducted on November 15, 2012 (UT) at BOAO (Bohyunsan Optical Astronomy Observatory) with an optical long-slit spectrograph installed on the Cassegrain focus of a 1.8-m telescope. The wavelength coverage was 5900 $-$ 7050 \AA~with 1200 gmm$^{-1}$ grism, which was chosen to achieve the emission lines of \Ha, \SII~$\lambda\lambda$6716,6731, \NII~$\lambda\lambda$6548,6584, \OI~$\lambda\lambda$6300,6363. The slit length and width were 3.6 arc-minute and 2.9 arc-second, respectively. The resultant spectral resolution was $\sim$ 2070 (150 \kms) with the dispersion of 0.41 \AA pixel$^{-1}$. The CCD camera size is 4k $\times$ 4k and the pixel scale along the slit is 0\farcs45 pixel$^{-1}$. Exposures on comparison lines were taken with the FeNeArHe lamp before and after the exposure of each object, for wavelength calibration.

Table \ref{tab:jkastable1} shows the observation log of HH 158 and HH 702. Figure \ref{fig:jkasfig1} shows the slit positions. The position angle (PA) of the HH 158 region including the central star is determined to be 223$\degree$ \citep{McGroarty2004} after verification through an HST image. In the case of HH 702 located at $\sim$ 11$^{\prime}$ from the star, a PA of 193$\degree$ was selected to obtain the emissions as much as possible. The exposure times were 1800 s and 2800 s for HH 158 and HH 702, respectively. HR 1544 (A1 type) was observed for telluric line correction and flux calibration.

We reduced the data with standard IRAF (Image Reduction and Analysis Facility) packages. Bias was subtracted from every frame. The flat-fielding was carried out using a normalized flat made by the APFLATTEN task. Bad pixels and cosmic-ray events were corrected. The spectra were extracted using the APALL task. The wavelength calibration and distortion correction were accomplished by the IDENTIFY, REIDENTIFY, FITCOORDS, TRANSFORM tasks. There were 1 and 2 pixel shifts between the comparison frames before and after the exposure for HH 158 and HH 702, respectively. Since the instrument is mounted on the Cassegrain focus, the instrumental displacement due to changes of the telescope elevation angle cause those shifts. We used the average of two lamp images to achieve a more accurate spectral calibration. There were also wavelength shifts between object frames. We corrected these shifts with the OH airglow lines in the object frames. After that, the sky emission lines were subtracted using the BACKGROUND task. The wavelength sensitivity correction and flux calibration were done with the standard star, HR 1544 (A1, V=4.35, T$_{\rm eff}$=9150 K).

\begin{figure*}[]
\centering
\includegraphics[width=170mm]{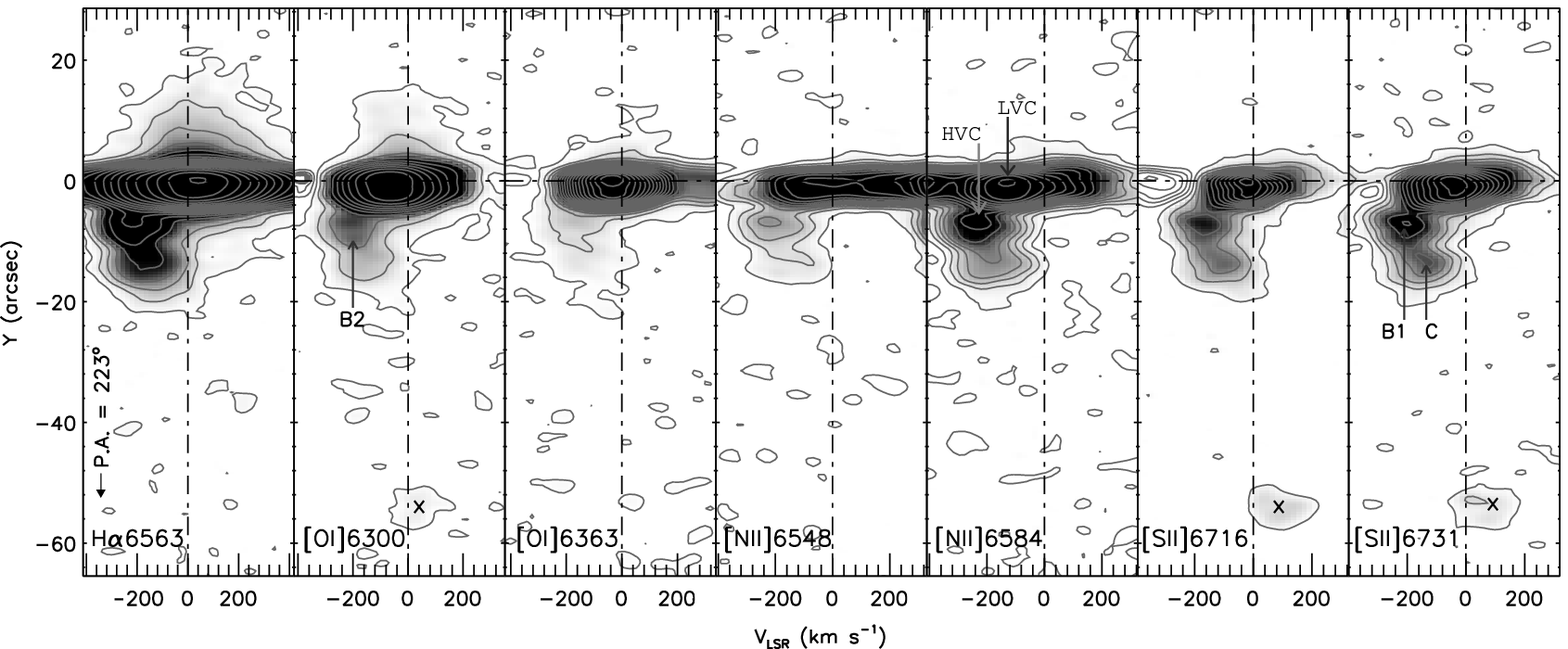}
\caption{Continuum-subtracted PVDs of the emission lines of HH 158. Contours are drawn from 0.001 (3 $\sigma$) to the peak levels of 5.2 (\Ha), 0.98 (\OI~$\lambda$6300), 0.33 (\OI~$\lambda$6363), 0.17 (\NII~$\lambda$6548 and $\lambda$6584), 0.10 (\SII~$\lambda$6716) and 0.22 (\SII~$\lambda$6731) $\times$ 10$^{-15}$ W m$^{-2}$ \AA$^{-1}$ with an equal interval in a logarithmic scale. The position angle (223$\degree$) of the slit is marked in the left panel. Northeast is up and southwest is down in the diagram. The original spectrum has Gaussian Smoothed with a $\sigma$ value of 1.5 $\times$ 1.5. Knots B1, B2 \citep{Lavalley1997}, and C \citep{Eisloffel1998} are identified at Y $\sim$ $-$6$\arcsec$ and $-$14$\arcsec$. The faint emissions at Y = $-$54$\arcsec$ in \OI~$\lambda$6300, \SII~$\lambda$6716 and \SII~$\lambda$6731 are parts of the HH 159 outflow ejected from DG Tau B. These are labeled with the `$\times$' symbols in the figure. \label{fig:jkasfig3}}
\end{figure*}

\begin{figure*}[]
\centering
\includegraphics[width=170mm]{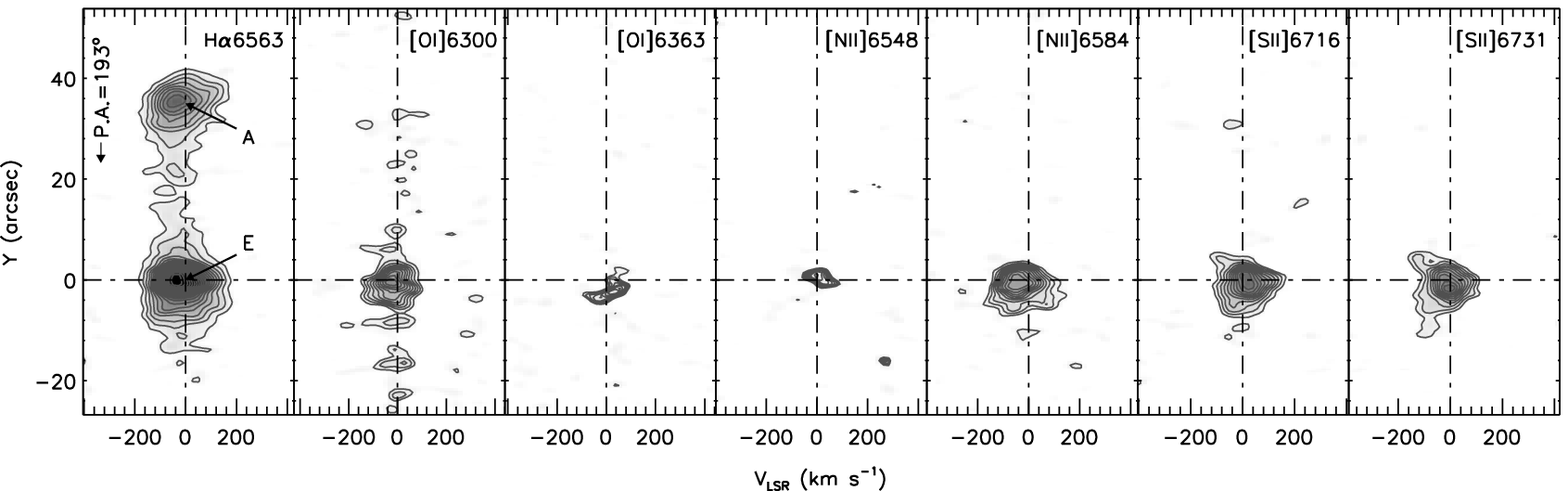}
\caption{PVDs of the emission lines of HH 702. Contours are drawn from 0.05 to the peak levels of 1.5 (\Ha), 0.45 (\OI~$\lambda$6300), 0.12 (\OI~$\lambda$6363), 0.11 (\NII~$\lambda$6548), 0.38 (\NII~$\lambda$6584), 0.60 (\SII~$\lambda$6716), and 0.58 (\SII~$\lambda$6731) in 10$^{-17}$ W m$^{-2}$ \AA$^{-1}$ with an equal interval in a logarithmic scale. The position angle (193$\degree$) of the slit is marked in the left panel. Northeast is up and southwest is down in the diagram. The knots A and E of \citet{McGroarty2004} are identified. \label{fig:jkasfig4}}
\end{figure*}

\begin{figure}[t!]
\centering
\includegraphics[width=83mm]{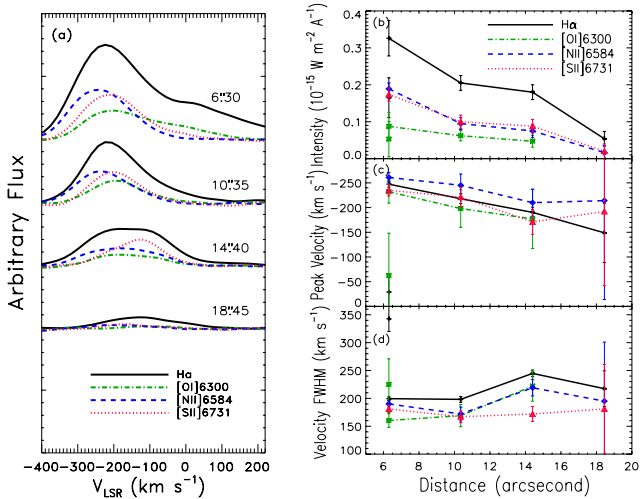}
\caption{($a$) Variation of the line profile, ($b$) peak intensity, ($c$) peak velocity, and ($d$) velocity width (FWHM) of the each emission line of HH 158 with the distance from DG Tau. \label{fig:jkasfig5}}
\end{figure}

We removed a deep \Ha~absorption line in the spectrum of the standard star using the SPLOT task before applying the wavelength sensitivity correction.

\section{Results\label{sec:result}}

\subsection{Kinematics}

Figure \ref{fig:jkasfig2} shows the 1D spectrum extracted from the central source region within $\pm$5\farcs4 from DG Tau. In the wavelength range of 6200 $-$ 6800\AA, the \Ha~$\lambda$6563, \OI~$\lambda\lambda$6300, 6363, \NII~$\lambda\lambda$6548, 6584, and \SII~$\lambda\lambda$6716, 6731 emissions were detected. The {\FeII}~emissions at $\lambda$6433, $\lambda$6456, $\lambda$6516 \AA~and the atomic Helium line at $\lambda$6678 \AA~are also identified. the \Ha~emission has the highest intensity peak level of 6 $\times$ 10$^{-14}$ W m$^{-2}$ \AA$^{-1}$, which is $\sim$ 30 times higher than that of the stellar continuum. The wing of the strong {\Ha}  line affects the line profiles of two {\NII}~lines.

Figures~\ref{fig:jkasfig3} and \ref{fig:jkasfig4} show the position-velocity diagrams (PVDs) of HH 158 and HH 702. In Figure~\ref{fig:jkasfig3}, the stellar continuum was subtracted. The position at Y = 0$\arcsec$ indicates DG Tau and the peak of knot E in Figures \ref{fig:jkasfig3} and \ref{fig:jkasfig4}, respectively. HH 702 is located at $\sim$ 11$^{\prime}$ ($=$ 0.45 pc at 140 pc distance) southwest from DG Tau, as shown in Figure \ref{fig:jkasfig1}. The Gaussian smoothing was applied in both figures with sigma of 1.5 $\times$ 1.5 pixels.

In Figure~\ref{fig:jkasfig3}, we identify the knots B1, B2 \citep{Lavalley1997} and C \citep{Eisloffel1998} at the distance of 6\farcs74, 8\farcs5 and $\sim$ 14$\arcsec$ from the star. The two velocity components appear in all emission lines in Figure \ref{fig:jkasfig3}. The low velocity component (LVC) is located within $\pm$5$\arcsec$. The peak velocity of the LVCs varies from $\sim$ $-$80 to $-$20 {\kms}. The high velocity component (HVC) is relevant at $|$Y$| >$ 6$\arcsec$. Figure~\ref{fig:jkasfig5}\textit{c} shows that the HVC is getting slower with increasing distance, e.g., $-$270 \kms~at 6{\farcs}3 (knot B1) and $-$150 {\kms}~at 18{\farcs}45. The real jet velocity is 1.18 times higher than the radial velocity considering the inclination angle of 32$\degree$ with respect to the line of sight \citep{Eisloffel1998}, so the radial velocity of $-$270 {\kms}~corresponds to the jet velocity of $-$320 {\kms}. Additionally, it is noticeable that the emission lines are showing different spatial extents. In Figure~\ref{fig:jkasfig3}, all emissions except {\NII} $\lambda$6548 show extensions to Y $\sim$ $-$20$\arcsec$ at 3 $\sigma$ contour levels. Knot B1 is relevant in all emissions except \OI~$\lambda$6363. Knot C is remarkable in \SII~$\lambda$6731 and appears as plateaus in {\Ha}, {\OI} $\lambda$6300, \NII~$\lambda$6584, and {\SII} $\lambda$6716. Its intensity in {\NII} $\lambda$6584 is almost twice than that of \OI~$\lambda$6300.

The faint emissions at a distance of 54$\arcsec$ from DG Tau, detected at \OI~$\lambda$6300, \SII~$\lambda$6716 and \SII~$\lambda$6731 may not come from HH 158 but from a part of HH 159 (see Figure~\ref{fig:jkasfig1}), which is ejected from DG Tau B. These features are labeled with `$\times$' symbols.

In Figure~\ref{fig:jkasfig4}, we identified the knots A and E of \citet{McGroarty2004}. The knot A of HH 702 is bright only in \Ha~emission, while it is seen in the {\SII} images of \citet{McGroarty2004} and \citet{Sun2003}. Knot E is detected in all emission lines.

\begin{figure}[t!]
\centering
\includegraphics[width=50mm]{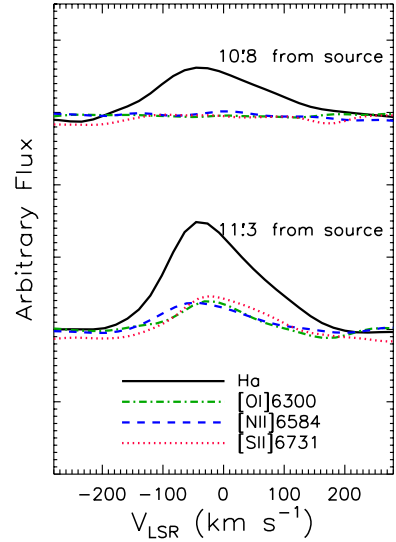}
\caption{Line profiles at knots A ($upper$) and E ($lower$) of HH 702.\label{fig:jkasfig6}}
\end{figure}

Figure~\ref{fig:jkasfig5} shows the variations of the line profile, peak intensity, peak velocity, and velocity width (FWHM) of each emission line with the distance from the star for HH 158. The data are sampled at 6{\farcs}30, 10{\farcs}35, 14{\farcs}40, and 18{\farcs}45 along the blueshifted jet. The plot starts from 6{\farcs}30 where the knot B1 is located. The \OI~$\lambda$6300 emission at 18{\farcs}45 could not be measured due to faintness. In Figures~\ref{fig:jkasfig5}\textit{b} and \ref{fig:jkasfig5}\textit{c}, the peak intensities and velocities of all the emission lines decrease with distance from the source. The uncertainty in the peak velocity measurement of {\NII} and {\SII} at 18{\farcs}45 is $\pm$170 \kms. It is large, due to low a signal-to-noise ratio (S/N $<$ 2). Even though  there are small increases in velocity (Figure~\ref{fig:jkasfig5}\textit{c}), those are not significant due to a large uncertainty. The decreasing rate in the velocity of {\Ha} is $|\Delta v_{peak}|$ $\sim$7 km/$\arcsec$.

\begin{figure*}[t!]
\centering
\includegraphics[width=140mm]{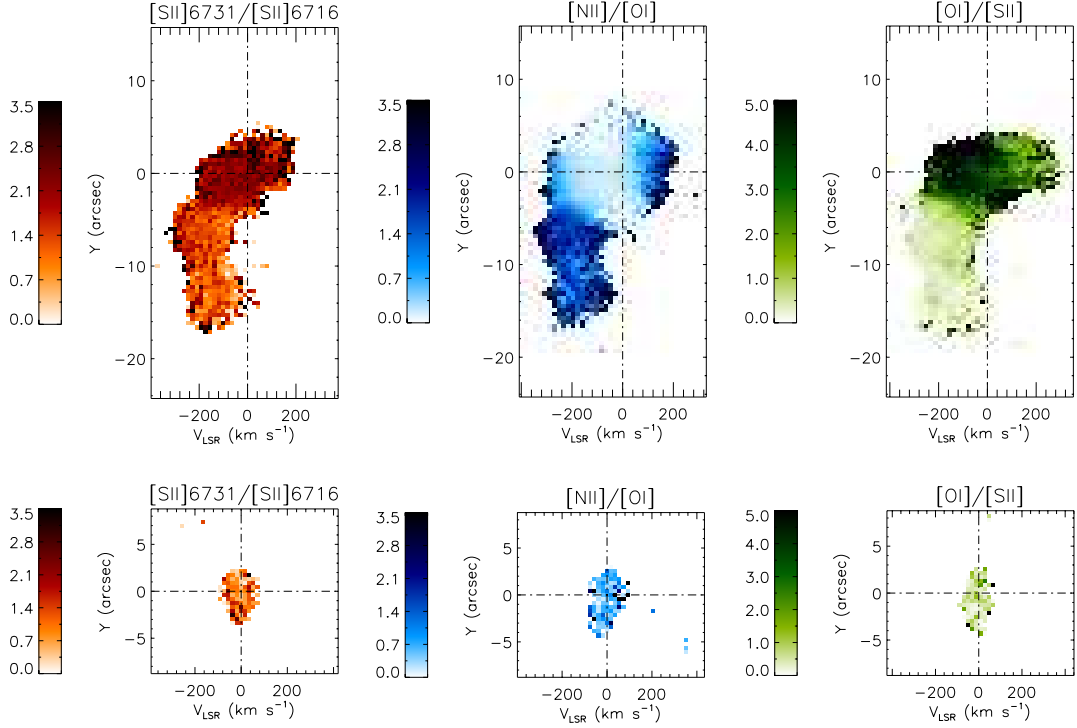}
\caption{Ratios of forbidden emission lines: (\textit{top}) HH 158 and (\textit{bottom}) knot E of HH 702. (\textit{Left}) {\SII} $\lambda$6731/ $\lambda$6716, (\textit{center}) {\NII}~$\lambda$6548+$\lambda$6584 / {\OI}~$\lambda$6300+$\lambda$6363, and (\textit{right}) {\OI}~$\lambda$6300+$\lambda$6363 / {\SII}~$\lambda$6716+$\lambda$6731. \label{fig:jkasfig7}}
\end{figure*}

In Figure~\ref{fig:jkasfig5}\textit{d}, the range of velocity widths (FWHM) varies from 150 to 350 {\kms}, which is marginally resolved considering the velocity resolution of 150 {\kms} in our observation. The low velocity component shows a wider width ($\sim$340 {\kms} in {\Ha}) than those estimated for high velocity components ($\sim$160 $-$ 230 {\kms}), compatible with results obtained in previous studies \citep[e.g.,][]{Hirth1997,Pyo2003}. \citet{Pyo2002} interpreted the wider velocity width of LVC as follows: if we assume that the outflow has the shape of a diverging wind, the velocity width represents the difference in radial velocities between the nearest and farthest streamlines from us. Thus, the wider velocity width means a wider opening angle.

Figure \ref{fig:jkasfig6} shows the variations of the line profile along the distance of HH 702.
The peak velocities of two knots are $\sim-$80 {\kms} in the {\Ha} emission.

\subsection{Line Ratios}

\begin{figure}[t!]
\centering
\includegraphics[width=50mm]{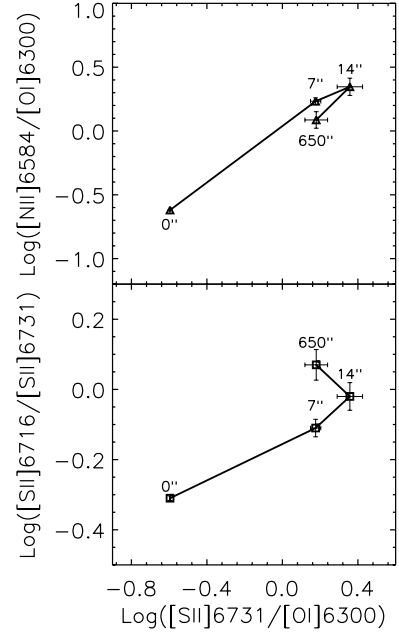}
\caption{Line ratios estimated from HH 158 and HH 702, for four distances: $d$ = 0$\arcsec$, 7$\arcsec$, 14$\arcsec$, 650$\arcsec$. \label{fig:jkasfig8}}
\end{figure}

\begin{table*}[t]
\caption{Locations of the knots with respect to the source position during the last 30 years.\label{tab:jkastable2}}
\centering
\begin{tabular}{lcccc}
\toprule
 & & Distance ($\arcsec$) &  & \\
\cline{2-4}
Observation date  & B1$^{\rm a}$ & B2$^{\rm a}$ & C$^{\rm b}$ & Reference  \\
\midrule
1984 Oct. 9 & -- & -- & 8.5 & \citet{Mundt1987}$^{\rm c}$ \\
1986 Dec.	& -- & -- & 8.7 &	\citet{Eisloffel1998}\\
1992 Oct.	& 2.25 & 3.25 & -- &\citet{Solf1993}\\
1994 Nov. 3 & 2.7 & 3.9  & -- & \citet{Lavalley1997}\\
1997 Jan.	&3.3 & -- & -- & \citet{Dougados2000}\\
1998 Jan. 23-26 &3.6 & -- & --&	\citet{Lavalley2000}\\
1998 Feb. 5	& -- & -- &10.65&	HST$^{\rm d}$\\
1999 Jan. 14&3.8& -- & -- &	\citet{Maurri2014}\\
2002 Sep. 17&4.6& -- &12	&\citet{Whelan2004}\\
2012 Nov. 15&6.74&8.75&13.7&	This work\\
2014 Nov. 21&7.65& -- &13.5&	This work\\
Proper motion ($^{\prime\prime}$yr$^{-1}$) &0.230& 0.272&0.180 & -- \\
Ejection date&1981	&1980&	1939& -- \\
\bottomrule
\end{tabular}
\tabnote{
$^{\rm a}$  knot name from \citet{Lavalley1997}.\\
$^{\rm b}$  knot name from \citet{Eisloffel1998}.\\
$^{\rm c}$  position of knot C measured from Figure 9 of \citet{Mundt1987}.\\
$^{\rm d}$  data taken from the HST archive, image of WFPC2 with the F675WF filter (proposal ID. 6855).\\
}
\end{table*}

The diagnostics by the ratios of forbidden emission lines is a very effective tool to study jet properties \citep{Hartigan1994, Bacciotti1995}. In Figure \ref{fig:jkasfig7}, we show the flux ratios of the forbidden emisson lines of HH 158 and knot E of HH 702 in PVDs : {\SII}~$\lambda$6731 / $\lambda$6716, {\NII}~$\lambda$6548+$\lambda$6584 / {\OI}~$\lambda$6300+$\lambda$6363, and {\OI}~$\lambda$6300+$\lambda$6363 / {\SII}~$\lambda$6716+$\lambda$6731.
The ratio of the \SII~doublet can be used to estimate the electron density \citep{Osterbrock1989}, which is larger than 2 within $\pm$5$\arcsec$ and decreases with the distance.
The {\NII}/{\OI}~ratio, which is proportional to the ionization fraction of hydrogen, is less than 0.5 at the center and increases at distances further than $\sim$ 5$\arcsec$. The {\OI}/{\SII}~ratio shows a similar trend with the {\SII}~doublet, large near the center and less than 1.0 further than 5$\arcsec$. This ratio is related to the electron temperature. The line ratios of HH 702 are comparable with those obtained from HH 158 at $\sim$ 14$\arcsec$ from the source.
In Figure \ref{fig:jkasfig8}, we show the relations between three ratios, {\SII}~$\lambda$6731 / {\OI}~$\lambda$6300, {\NII}~$\lambda$6584 / {\OI}~$\lambda$6300, and {\SII}~$\lambda$6716 / $\lambda$6731. In the figure, the four points represent the values at distances $d$ = 0$\arcsec$, 7$\arcsec$, 14$\arcsec$, 650$\arcsec$ from the source. The distance of 650$\arcsec$ is the location of the knot E of HH 702. The ratios increase with distance, as in a previous study \citep[Figure 3a--b]{Lavalley2000}.

\section{Discussion\label{sec:discussion}}

\subsection{Velocity Variation}

In Figure~\ref{fig:jkasfig3}, the HVC is getting slower with distance. The velocity decreasing rate derived in Section \ref{sec:result} is $\sim$ 7 km/$\arcsec$. If the jet is continuously slowing, then it should stop at 35$\arcsec$ from DG Tau. However, the knots of HH 702 located at 650$\arcsec$ show a radial velocity of $\sim$ $-$80 {\kms}.

Radial velocity ($v_{rad}$) measurements of knot C in HH 158 were reported in several previous studies. The knot was observed at 8\farcs5 \citep{Mundt1987} and 8\farcs7 \citep{Eisloffel1998} from the source in 1984 and 1987, respectively. We converted the heliocentric velocities to LSR velocities for the comparison with our data. In \citet{Mundt1987}, $v_{rad}$ were $-$136 {\kms} in {\Ha} and $-$123 {\kms} in {\SII} $\lambda$6731. In \citet{Eisloffel1998}, {\SII} is $\sim$ $-$146 {\kms}. In 2002 \citep{Whelan2004}, {\SII} $\lambda$6731 and {\OI} $\lambda$6300 show $v_{rad}$ of $\sim$ $-$150 and $-$200 {\kms}, respectively.
For the three emission lines mentioned above ({\Ha}, {\SII}, {\OI}), our data show $v_{rad}$ of $-$190, $-$128, and $-$180 {\kms}, respectively. Since different emission lines originate from different excitation conditions (i.e., temperature, density, etc.), they represent various regions in the outflow and show peaks in different velocities. The temporal variation of the velocity indicates that the velocity of knot is variable not only along the distance but also in time.
This may imply that the ejected knot has been decelerated when it passes through the ambient medium. The velocity change of the knot could also occur due to the shock caused by the interaction between the faster gas stream behind and the slower front knot.

\begin{figure}[t!]
\centering
\includegraphics[width=80mm]{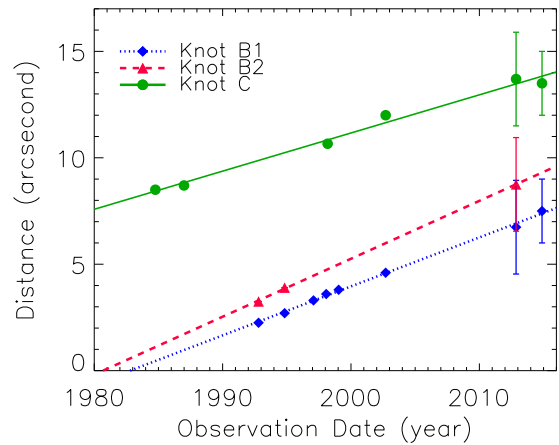}
\caption{Positional variations of three knots. For knot B1, the data points are expanded from those in \citet{Pyo2003}. \label{fig:jkasfig9}}
\end{figure}

\subsection{Proper Motion}

In the emission lines of HH 158, we have distinguished three knots at 6{\farcs}74 and 8{\farcs}75, and 13{\farcs}7 from the source in our data taken on November 15, 2012.
We identified these knots as B1 and B2 of \citet{Lavalley1997}, and knot C of \citet{Eisloffel1998}.
Recently, we could measure the knot positions from our additional data taken on November 21, 2014. For knot B1 and C, it was measured to 7{\farcs65} and 13{\farcs}5, respectively.
We note that knot B2 is not resolved in our spectra in 2014. The distance of knot C from the source is 0{\farcs}2 smaller in 2014 than that  in 2012. The seeing size is $>$ 2$\arcsec$, so it may be due to the uncertainty of the measurement.

\begin{figure}[]
\centering
\includegraphics[width=80mm]{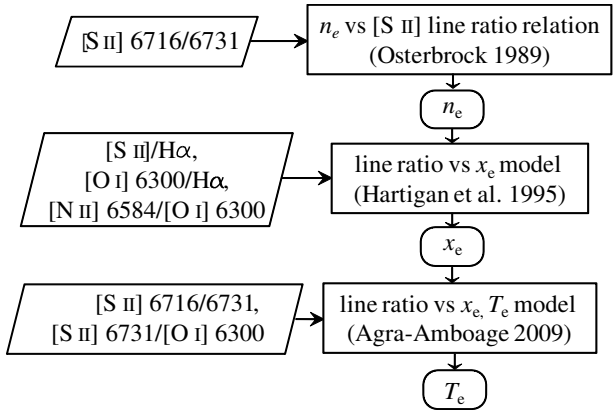}
\caption{Flow chart showing how we derive the electron density ($n_{e}$), ionization fraction ($x_{e}$), and electron temperature ($T_e$). \label{fig:jkasfig10}}
\end{figure}

In Table \ref{tab:jkastable2}, we list the locations of the knots during the last $\sim$ 30 years including our work. Basically, the positions of™" knot B1 are extended from Table 1 of \citet{Pyo2003}.
In the case of knot C, the first detection was reported in 1984 \citep{Mundt1987} and it was seen at 12$\arcsec$ in long-slit spectroscopy of \citet{Whelan2004} whose data were taken in 2002.
In addition to these data points, we found an Hubble Space Telescope (HST) archival image obtained in 1998. The central peak in intensity contour of the knot in the HST image is located about 10{\farcs}65 from DG Tau.

\begin{figure*}[]
\centering
\includegraphics[width=150mm]{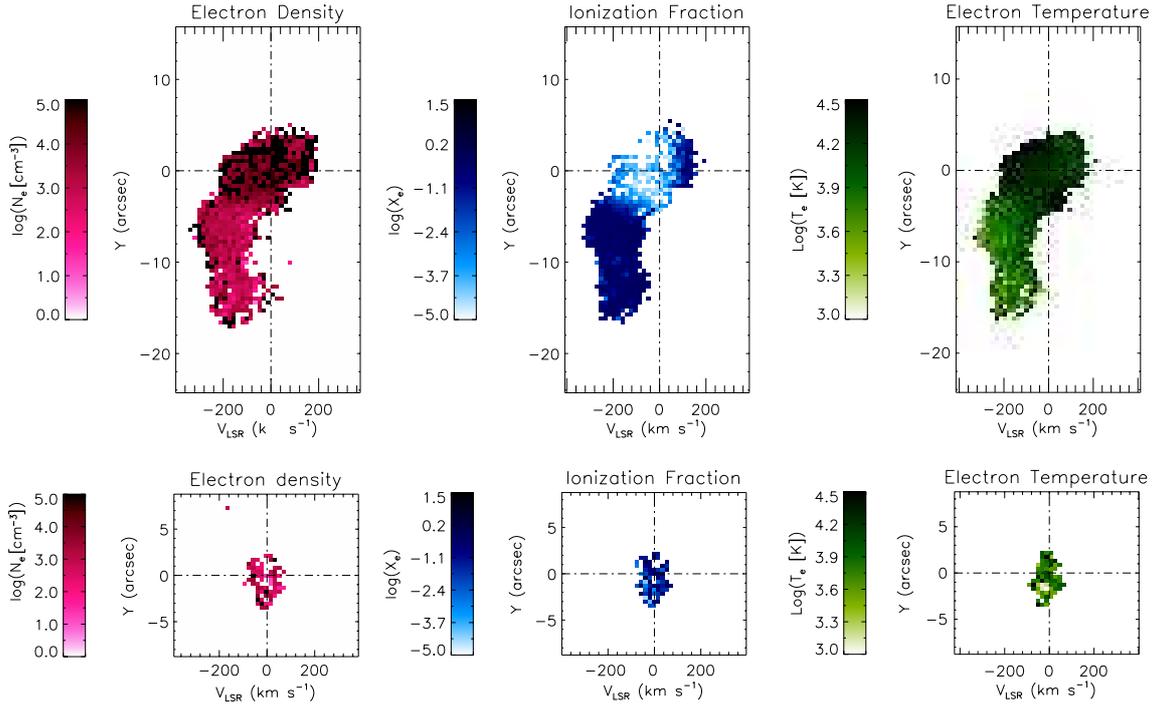}
\caption{ (\textit{Left}) Electron density ($n_{e}$), (\textit{center}) ionization fraction ($x_{e}$), and  (\textit{right}) electron temperature ($T_e$) of HH 158 and HH 702. \label{fig:jkasfig11}}
\end{figure*}

Figure~\ref{fig:jkasfig9} shows the temporal variation of the distance for the knots B1, B2, and C.
We estimated the proper motion by linear regression: 0{\farcs}230 $\pm$ 0{\farcs}006 yr$^{-1}$ for B1, 0{\farcs}272 $\pm$ 0{\farcs}002 yr$^{-1}$ for B2, 0{\farcs}180 $\pm$ 0{\farcs}008 yr$^{-1}$ for C (see Table~\ref{tab:jkastable2}). For knot B1, the estimated value is smaller than those obtained from previous studies, e.g., 0{\farcs}272 in \citet{Pyo2003}. Due to the large spatial resolution of our observation, the emission from knot B1 could be combined with that from knot A1 which has identified from previous studies \citep{Bacciotti2000, Dougados2000}. The smaller proper motion value of knot B1 may be estimated because knot A1 is located $~$2{\farcs}5 closer \citep{Maurri2014} to the star than knot B1.
From the radial and the tangential velocities of each knots, the inclination angles ($i$) of
B1, B2, and C are calculated as 32\fdg5 $\pm$ 0\fdg7, 37\fdg0 $\pm$ 0\fdg2, 32\fdg3 $\pm$ 1\fdg1 with respect to the line of sight.

For HH 702, \citet{McGroarty2007} estimated the tangential velocities of knots A and E as 129 $\pm$ (10 $-$ 28) and 186 $\pm$ (10 $-$ 28) {\kms} in the {\Ha} emission, which correspond to 0{\farcs}19 $\pm$ 0{\farcs}03 and 0{\farcs}28 $\pm$ 0{\farcs}03 yr$^{-1}$ in proper motion, respectively. From the radial velocity of $\sim$ $-$80 {\kms} and the tangential velocities, the $i$ of knot A and E are estimated as $\sim$58\fdg2 $\pm$ 1\fdg1 and $\sim$66\fdg7 $\pm$ 1\fdg1 with respect to the line of sight. The $i$ of HH 702 are $\sim$21$-$35$\degree$ different from those of HH 158. We cannot avoid the possibility that the two HH objects were ejected from different sources although \citet{McGroarty2007} postulated that DG Tau seems the most probable driving source of HH 702 based on the proper motion direction. The different direction or the curved shape of a jet may indicate the precession of the jet \citep{Raga2001}.

\subsection{Line Ratios and Physical Parameters}

\citet{Maurri2014} shows the line ratios and physical parameters within 5$\arcsec$ of DG Tau. Their \SII~$\lambda$6731/$\lambda$6716 ratio is similar to our value. Knots B0 and B1 at 3{\farcs}3 and 3{\farcs}8 in \citet{Maurri2014} correspond to B1 at 6{\farcs}7 in our data, after considering the proper motion. For this knot, the {\SII}~doublet ratio in our data is slightly smaller than their estimation. For the {\NII}/{\OI}~ratio, the value in \citet{Maurri2014} is 0.01 $-$ 2.5, in a similar range as our data.

In Figure \ref{fig:jkasfig10}, the flow chart showing how we estimate the electron density ($n_{e}$), ionization fraction ($x_{e}$), and electron temperature ($T_{e}$) is displayed. Figure \ref{fig:jkasfig11} shows the derived physical parameters for HH 158 and HH 702. The estimated properties of each knot show good agreement with the planar shock model \citep{Hartigan1994,Lavalley2000}. For the variation of the jet with the distance (i.e., variation with time), the model that treats the time variability of the jet \citep{Raga1990,Raga1992} could explain our result with the internal shock caused by a variable (or periodic) jet ejection.

In our data, it is hard to distinguish the two velocity components within $\pm$ 5$\arcsec$ from DG Tau due to the limitations in spatial and velocity resolution. The LVC is usually dominant close to the source, while HVC is extended further as shown in Figure \ref{fig:jkasfig3}. With this point of view, it is noticeable that the $x_{e}$ of HVC at $|$Y$| >$ 4$\arcsec$ is higher than that of the LVC at $|$Y$| <$ 4$\arcsec$ but $n_{e}$ is lower in HVC in Figure \ref{fig:jkasfig11}. One the other hand, previous studies \citep{Lavalley2000, Maurri2014} show that the $n_{e}$ and $x_{e}$ are larger in their high and medium velocity interval than those in low velocity. \citet{Hamann1994} also showed that the two velocity components of 20 TTS and 12 HAeBe stars are different in $T_{e}$ and $n_{e}$. These results indicate that the two velocity components have different physical conditions. They are supposed to originate from different parts of star-disk systems or have different launching mechanisms \citep{Kwan1995, Pyo2003, Pyo2006}.

\subsubsection{Electron density ($n_{e}$)}
\begin{figure}[]
\centering
\includegraphics[width=70mm]{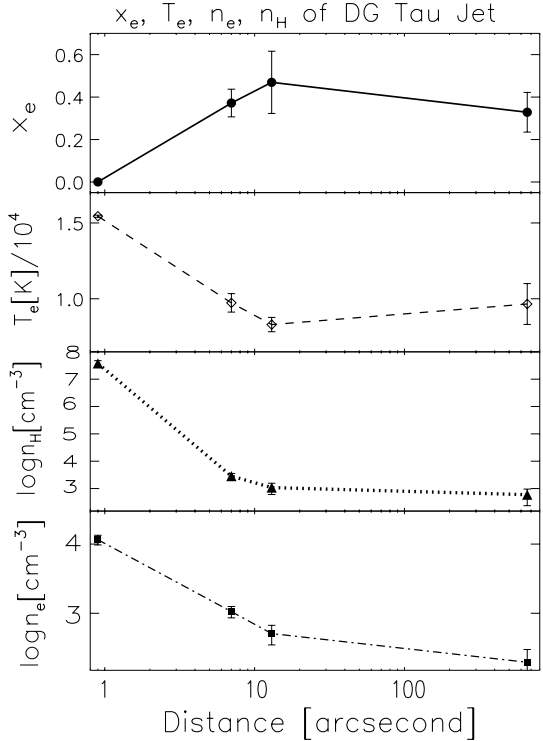}
\caption{ (\textit{From top to bottom}) Ionization fraction ($x_{e}$), \textit{}electron temperature ($T_{e}$), total hydrogen density ($n_{H}$), and electron density ($n_{e}$) of DG Tau jet as a function of the distance from the source. \label{fig:jkasfig12}}
\end{figure}

We calculated the electron density using the typical method of {\SII}~$\lambda$6716/$\lambda$6731 doublet ratio \citep{Osterbrock1989}. This ratio is known to be insensitive to temperature variation, so we assumed the electron temperature as $\sim$ 10,000 K to estimate the electron density ($n_{e}$).
In the bottom panel of Figure \ref{fig:jkasfig12}, $n_{e}$ is shown at four different distance positions. At the proximity of the star, $n_{e}$ is $\sim$ 10$^{4}$ cm$^{-3}$ and decreases to $\sim$ 500 cm$^{-3}$ at 14$\arcsec$ from the source. In HH 702 at 650$\arcsec$, $n_{e}$ is estimated to be $\sim$ 200 cm$^{-3}$.
\subsubsection{Ionization fraction ($x_{e}$)}\label{sec:ionization}

\begin{figure}[]
\centering
\includegraphics[width=83mm]{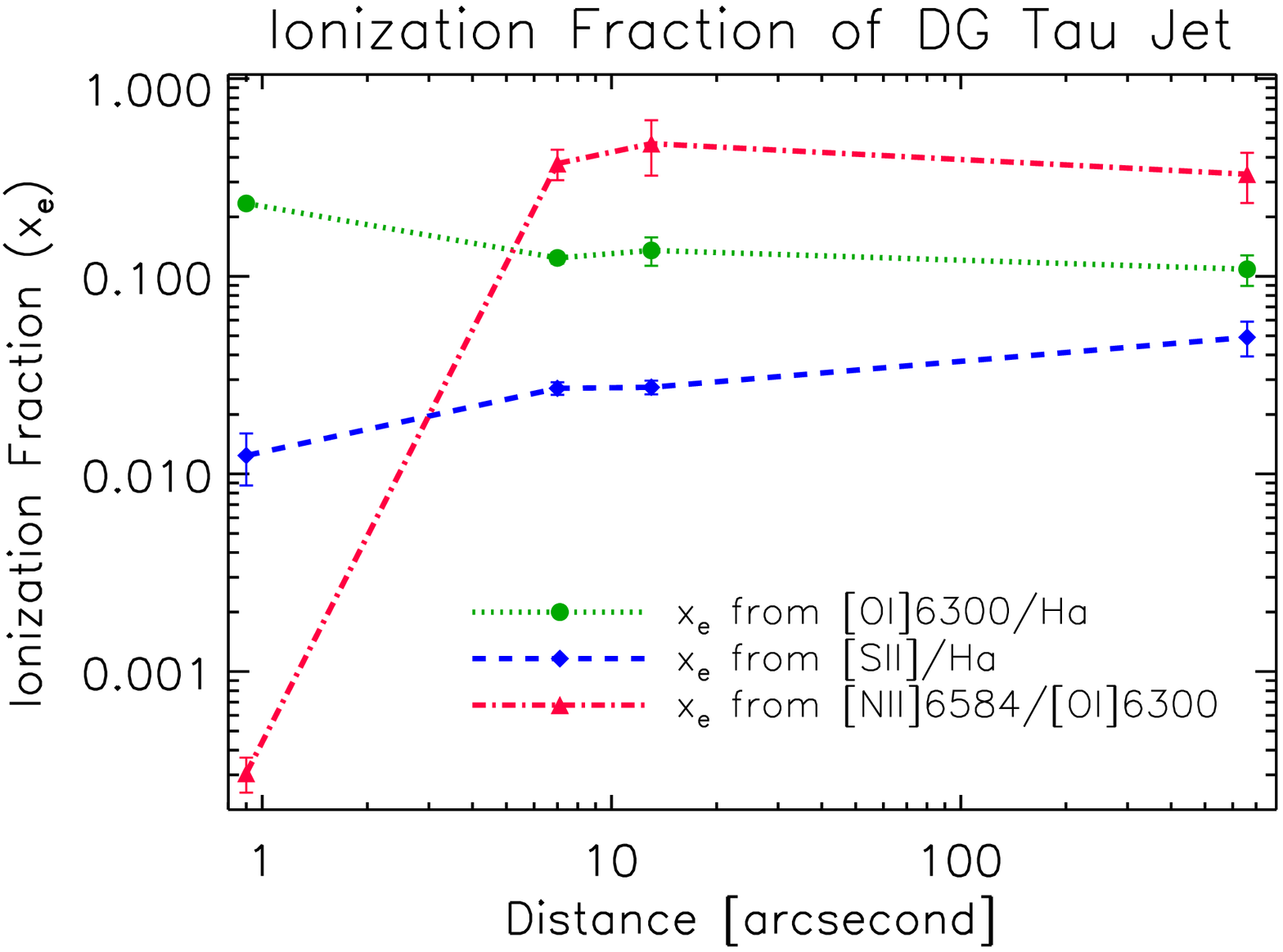}
\caption{Ionization fraction ($x_{e}$) derived from the line ratios of {\OI}/{\Ha}, {\SII}/{\Ha}, and {\NII}/{\OI} based on the model of \citet{Hartigan1994}. \label{fig:jkasfig13}}
\end{figure}

We estimated the ionization fraction from the model grid of \citet{Hartigan1994}, which shows the relations between the ionization fraction and the line ratio of \OI~$\lambda$6300/{\Ha}, {\SII}~$\lambda$6716+$\lambda$6731/{\Ha}, {\NII}~$\lambda$6548/{\OI}~$\lambda$6300 in Figures 3, 4, and 6. We assumed the magnetic strength of 100 $\mu$G, close to the average value in the model.
The middle panel of Figure \ref{fig:jkasfig11} and the top panel of \ref{fig:jkasfig12} show the ionization fraction obtained from the {\NII}/{\OI}~ratio. Figure \ref{fig:jkasfig13} shows the $x_{e}$ values derived from three different ratios of {\OI}~$\lambda$6300/{\Ha}, {\SII}~$\lambda$6716+$\lambda$6731/{\Ha}, {\NII}~$\lambda$6548/{\OI}~$\lambda$6300.
The $x_{e}$ values deduced by {\OI}/{\Ha}~and {\SII}/{\Ha}~ratios are in the range of 0.02 $-$ 0.2 which are much different from the values deduced from {\NII}/{\OI}.
Since the {\Ha}~emission flux comes not only from hydrogen recombination but also from collision \citep{Bacciotti1999, Hartigan2003}, the use of the \Ha~emission in the estimation of the ionization fraction may cause uncertainty. Thus, we used only the {\NII}/{\OI}~ratio in the calculation of $x_{e}$.

In Figures \ref{fig:jkasfig11} and \ref{fig:jkasfig12}, $x_{e}$ varies from $\sim$ 10$^{-3}$ to $\sim$ 0.5 with distance from the source. According to Figure~16 of \citet{Hartigan1994} which shows the relation between ionization fraction and shock velocity, the shock velocity surpasses 100 {\kms}~with $x_{e}$ of $\sim$ 0.5 around knot C, which has $\sim$190 {\kms} radial velocity. \citet{Lavalley2000} showed that $x_{e}$ was increasing from 0.02 to 0.6 in HVC and from 0.004 to  0.1 in LVC within 4$\arcsec$ from the source. Recently, \citet{Maurri2014} reported similar results within 5$\arcsec$ using HST/STIS data obtained in 1999.

For HH 702, $x_{e}$ is greater than 0.2, which is similar to the value at 14$\arcsec$ from the source.

\subsubsection{Electron temperature ($T_{e}$)}

Electron temperature ($T_{e}$) values are shown in Figures \ref{fig:jkasfig11} and \ref{fig:jkasfig12}, and are estimated from the model grid of Figure 3.6 in \citet{Agra-Amboage2009} which shows the relation between emission line ratios and electron density, temperature, and ionization fraction. Using the $x_{e}$ values obtained from \ref{sec:ionization}, we have calculated the temperature from the model. $T_e$ is higher than 15,000 K in the central region and decreases to $\sim$ 5,000 K at 14$\arcsec$. At HH 702, the temperature was estimated to be slightly higher than that measured at knot C of HH 158.

\subsection{Mass-loss Rate}

The mass-loss rate of DG Tau was estimated in several previous studies. \citet{Hartigan1995} estimated the value of 3.2 $\times$ 10$^{-7}$ $M_{\odot}$ yr$^{-1}$ from the \OI~$\lambda$6300 emission line flux. In \citet{Lavalley2000}, they have obtained the total value of 1.4 $\times$ 10$^{-8}$ $M_{\odot}$ yr$^{-1}$ from the low, intermediate, and high velocity interval. The lower limit of mass-loss rate estimated by \citet{Beck2010} using the high-velocity Br$\gamma$ emission was 1.2 $\times$ 10$^{-8}$ $M_{\odot}$ yr$^{-1}$. \citet{Agra-Amboage2011} has estimated a total mass-loss rate from $-$300 \kms~to $-$50 \kms~components of the \FeII~emission to 3.3 $\times$ 10$^{-8}$ $M_{\odot}$ yr$^{-1}$. \citet{Maurri2014} also has estimated a mass-loss rate in the range of 10$^{-8}$ to 10$^{-9}$  $M_{\odot}$ yr$^{-1}$ within the region 0\farcs7 from the source.

To estimate the mass-loss rate, we have adopted the method of \citet{Hartigan1995}. We can calculate the mass-loss rate with the equation below:
\begin{equation}
\label{eq:mass-loss}
\dot{M}_J = M_{TOT} V_{\bot} / l_{\bot}
\end{equation}
$V_{\bot}$ is the projected velocity on the plane of the sky and $l_{\bot}$ is the projected jet size of the aperture. The total mass of the jet M$_{TOT}$ is estimated from the forbidden {\OI} emission line with the following equation:
\begin{equation}
M_{TOT} = 9.61 \times 10^{-6} \left(1+\frac{n_c}{n_e}\right) \left(\frac{L_{6300}}{L_{\odot}}\right) M_{\odot}
\end{equation}

We estimate the mass-loss rate close to the star in the region within $\pm$ 5\farcs4 ($\sim$750 AU) from the star. The equivalent width of \OI~$\lambda$6300 emission line is measured to be ~21 {\AA}. We assume $R =$ 8.74 magnitude, which is the dereddened R magnitude of DG Tau in \citet{Hartigan1995}. The line luminosity of the \OI~$\lambda$6300 ($L_{6300}$) converted into solar units is 9.2 $\times$ 10$^{-3}$ $L_{\odot}$. $V_{\bot}$ is estimated to be 75 {\kms} for the measured radial velocity 120 {\kms}, applying the inclination angle of 32$\degree$ with respect to the line of sight. For $l_{\bot}$, we adopted 6.07 $\times$ 10$^{10}$ km as the projected jet aperture size which corresponds to our slit-width of 2{\farcs}9 at a 140 pc distance to the Taurus molecular cloud. The FWHM of the jet measured from \citet{Dougados2000} is $\sim$0{\farcs}21$-$1{\farcs}25 at a distance of $\sim$0{\farcs}4$-$4$\arcsec$ from the star. \citet{Agra-Amboage2011} obtained the value of $\sim$0{\farcs}1$-$0{\farcs}8 within $\sim$1{\farcs}4 from the source, for the medium velocity component of the blueshifted jet. The jet width increases with distance in the vicinity of the source. We assume that the jet width in our estimation would be larger than 1{\farcs}25 because we estimate a mass-loss rate within $\pm$ 5{\farcs}4. $l_{\bot}$ of 2{\farcs}9 should be regarded as an upper limit. The temperature is assumed to be 8200 K, the average temperature for the \OI~$\lambda$6300 line \citep{Hartigan1994} and the critical density $n_{c}$ in this temperature is 1.97 $\times$ 10$^{6}$ cm$^{-3}$. For the electron density $n_{e}$, 7 $\times$ 10$^{4}$ cm$^{-3}$ is used. The mass-loss rate, $\dot{M}_J$, for the {\OI} emission line is $\sim$ 1.0 $\times$ 10$^{-7}$ $M_{\odot}$ yr$^{-1}$.

The $\dot{M}_J$~can be also estimated using the {\SII}~$\lambda$6731 emission line. $M_{TOT}$  is given by:
\begin{equation}
M_{TOT} = 1.43 \times 10^{-3} \left(\frac{L_{6731}}{L_{\odot}}\right) M_{\odot}
\end{equation}
The equivalent width of the \SII~$\lambda$6731 line is $\sim$ 4.5 {\AA}.
The $\dot{M}_J$ from {\SII} is ~1.1 $\times$ 10$^{-7}$ $M_{\odot}$ yr$^{-1}$, which is almost the same as that of \OI~$\lambda$6300. The mass-loss rates of \OI~$\lambda$6300 and \SII~$\lambda$6731 lines are a factor of 3 lower than the value of \citet{Hartigan1995} and is 3 $-$ 7 times larger than those estimated by \citet{Lavalley2000} and \citet{Agra-Amboage2011}. The calculated rate can differ with time because the measured line intensities and velocities of the jet could change due to the variability of the accretion/outflow rate of YSO. The rate becomes about two times larger if we replace the jet aperture size in our calculation with the value in \citet{Hartigan1995}, which is 1{\farcs}5. The difference in spatial resolution between ours and higher-resolution \citep[e.g.,][]{Agra-Amboage2011} observations cloud cause the difference in the mass-loss rate.

\section{Summary\label{sec:summary}}

We have investigated the kinematics and physical characteristics of HH 158 and HH 702 from the optical long slit spectroscopy. From the analysis through the PVDs and line profiles depending on the distance from the source, the peak velocity varies in the range of $-$50 to $-$250 {\kms} and the line width was measured to 150 to 250 \kms.
The proper motions were measured for knot B1, B2 and C of HH 158:  0{\farcs}230, 0{\farcs}272, and 0{\farcs}180 yr$^{-1}$ for the knots, respectively.

From our analysis of the line ratios, we find that the physical properties ($n_e$, $x_e$, $T_e$) are changing along the jet from the proximity to the star to knot C at 14$\arcsec$: (1) $n_e$ varies from an order of 10$^{4}$ to 10$^{2}$ cm$^{-3}$, (2) $x_e$ varies from $\sim$ 0 to 0.4 at 14$\arcsec$, (3) $T_e$ varies from 15,000 K to $\sim$ 5,000 K.
It is noticeable that the line ratios and physical parameters of HH 702 do not show much difference from those obtained at knot C of HH 158.
The mass-loss rate ($\dot{M}_J$) is estimated to be $\sim$ 10$^{-7}$ $M_{\odot}$ yr$^{-1}$, similar to values obtained in previous studies.


\acknowledgments

We thank the staff of the BOAO (Bohyunsan Optical Astronomy Observatory) for their support during the observing runs.



\end{document}